# Deployment and Evaluation of an EHR-integrated, Large Language Model-Powered Tool to Triage Surgical Patients

Jane Wang, MD[1,2*]; Timothy Keyes, PhD[2,3,4*]; April S Liang, MD[1,2]; Stephen P Ma, MD, PhD[1,2]; Jason Shen[2]; Jerry Liu, MD[1,2]; Nerissa Ambers, MPH[5]; Abby Pandya, MBA, MS[3]; Rita Pandya, MD[1,2]; Jason Hom, MD[1,2]; Natasha Steele, MD, MPH[1,2**]; Jonathan H Chen, MD, PhD[1,2,4**]; Kevin Schulman, MD[1,2,4**]

*Co-first authors
**Co-senior authors

1. Division of Hospital Medicine, Department of Medicine, Stanford University, Palo Alto, CA;
2. Stanford University School of Medicine, Palo Alto, CA
3. Stanford Health Care, Technology and Digital Solutions, Palo Alto, CA
4. Division of Computational Medicine, Stanford University, Palo Alto, CA
5. Stanford Health Care, Nursing Informatics, Palo Alto, CA

Corresponding Author:
Jane Wang, MD
300 Pasteur Dr.
Palo Alto, CA 94304
Jwang007@stanford.edu
(510) 896-9842

# Abstract

*Importance:*
Large language models (LLM) are increasingly prevalent in clinical workflows. It is important to understand how these tools are built effectively and deployed safely.

*Objective*:
Surgical co-management (SCM) is an evidence-based care model in which hospitalists jointly manage medically complex perioperative patients alongside surgical teams. Despite its clinical and financial value, effective use of SCM is limited by the need to manually identify eligible patients. Our objective was to determine whether SCM eligibility triage can be automated.

*Design*:
This work describes a prospective, unblinded study at Stanford Health Care from September 2025 to February 2026. An LLM-based, electronic health record (EHR)-integrated, human-in-the-loop surgical triage tool (SCM Navigator) provided SCM triage recommendations, followed by physician review.

*Participants*:
All SCM attending physicians were invited to participate; 14 of 16 participated.

*Intervention*:
Using pre-operative documentation, structured data, and clinical criteria for peri-operative morbidity, SCM Navigator categorized patients as appropriate, not appropriate, or possibly appropriate for SCM. SCM faculty indicated clinical judgment of SCM appropriateness and provided free-text feedback when they disagreed with SCM Navigator.

*Main Outcomes and Measures:*
The sensitivity, specificity, positive predictive value (PPV), and negative predictive value (NPV) of SCM Navigator were measured using physician feedback as a reference. Free-text reasons were thematically categorized. Manual chart review was conducted on all false-negative cases. For the largest false-positive category, manual chart reviews were conducted on 15 randomly selected cases subsequently seen by SCM and 15 randomly selected cases not seen by SCM.

*Results:*
Since deployment, 6,193 cases have been triaged, of which 1,582 (23%) were recommended for hospitalist consultation. Using treating physicians' final determinations as the reference standard, SCM Navigator displayed high sensitivity (0.94, 95% CI 0.91–0.96) and moderate specificity (0.74, 95% CI 0.71-0.77) for identifying patients appropriate for SCM. Post-hoc chart review suggested that most discrepancies reflect modifiable gaps in clinical criteria, institutional workflow, or physician practice variability, rather than LLM misclassification, which accounted for 2 of 19 (11%) false-negative cases.

*Conclusions and Relevance:*
An LLM-powered, EHR-integrated, human-in-the-loop AI system can accurately and safely triage surgical patients for SCM. AI-enabled screening tools can augment and potentially automate time-intensive clinical workflows.

# Introduction

Artificial intelligence (AI) is entering clinical workflows in numerous capacities, from ambient documentation[1–3] to clinical decision support[4–6] and revenue cycle automation[7,8]. AI may be especially disruptive in healthcare because large language models (LLMs) can operate directly on clinical language to support high-volume cognitive work with potential benefits for operational efficiency and workforce burden[9,10]. Real-world deployments remain limited, with most studies conducted outside of live clinical workflows, without real patient data, and under inconsistent definitions of "readiness" for clinical use[11,12]. Despite growing enthusiasm, a translational chasm remains between AI's capabilities in research settings and evidence of safety and efficacy in real-world care[13].

Surgical co-management (SCM) is a rapidly growing care model that reduces clinical complications, inpatient length of stay, and health system costs[14–19]. SCM hospitalists collaborate with surgical and anesthesia services to care for perioperative patients with complex medical co-morbidities[20,21] by managing chronic conditions and complications perioperatively. SCM workflows at many institutions rely on manual identification of patients, leading to high inter-provider variability while contributing to physician burnout[22], reduced patient satisfaction[23], and worse clinical outcomes[24].

SCM screening serves as a natural early candidate for automation: the task is protocolized, the data is readily available, and the structured decision-making required is easily audited. Accordingly, we designed, built, deployed, and prospectively evaluated an EHR-integrated, LLM-powered triage tool (SCM Navigator) to automate SCM-eligibility assessment.

# Objective

To determine whether an EHR-integrated, LLM-powered agentic workflow can automate surgical case triage for SCM consultation.

# Methods

## Setting and Study Overview

The study population comprised surgical cases in Neurosurgery, Orthopedics, and Otolaryngology—departments currently covered by SCM—at Stanford Health Care's (SHC) Palo Alto campus from December 2024 to February 2026. SCM hospitalists care only for floor-level patients. Patients managed by the Palo Alto Medical Foundation (PAMF)—an outside hospitalist group—are excluded from SCM coverage and require manual identification. A detailed SCM workflow is illustrated in **Supplemental Figure 1**.

This study included two phases: (1) retrospective tool development/validation (December 2024-August 2025) and (2) prospective deployment/evaluation in clinical operations (September 2025 - February 2026).

## Technology Development and Integration

SCM Navigator is a human-in-the-loop, LLM-powered agentic workflow developed using ChatEHR, SHC's proprietary platform for securely deploying LLMs with direct access to EHR data.[25] Unlike fully autonomous agents that direct their own control flow, SCM Navigator uses pre-defined code paths to orchestrate the LLM: EHR inputs are assembled programmatically, the triage rubric is applied via a fixed prompting strategy, and outputs are post-processed into a structured format for EHR integration.[26] This design maximizes predictability, auditability, and operational reliability while reducing variance in runtime behavior, complexity, and cost that accompany higher-autonomy agentic systems.[27] A conceptual diagram of SCM Navigator is provided in **Figure 1a**.

## Clinical eligibility criteria and prompt design

A structured triage rubric was developed to operationalize clinical factors associated with perioperative morbidity based on both published literature[28,29] and SCM faculty clinical experience. Criteria were translated into a prompt instructing the LLM to triage each case for SCM eligibility and to provide an explanation with evidence from the chart. The triage criteria are described in **Table 1**. SCM Navigator's classification prompt (Appendix 1) was fixed before deployment and evaluation.

## Agentic workflow overview

For each scheduled case, SCM Navigator queries the patient's EHR and applies the triage rubric to the pre-operative anesthesia evaluation note and active medication list. It uses a three-tier output: "Affirmative" functions as a high-priority (strong) flag, "Maybe" as a lower-priority (weak) flag for borderline cases, and "Negative" indicates that no consultation is recommended. For each patient, it produces a single output string containing the classification (Affirmative/Maybe/Negative) and a brief explanation with supporting evidence. To separate clinical reasoning from output formatting/schema compliance, a second LLM call converts the string into a structured JSON object for downstream parsing. SCM Navigator then writes the classification and the explanation as Smart Data Elements into the Epic chart, where they can be reviewed by clinicians as a structured report (**Figure 1b**).

Beyond medical complexity criteria, SCM Navigator incorporates structured EHR fields to assign "Negative" classifications to cases (1) scheduled as an outpatient procedure without planned admission/overnight observation, (2) scheduled at a location outside SHC's main hospital, or (3) managed by an external provider group (the Palo Alto Medical Foundation; PAMF).

SCM Navigator analyzes the following day's surgical cases, presenting recommendations for SCM physician review. SCM physicians indicate their clinical assessment by selecting "Yes," (anticipate co-managing), or "No," (do not anticipate co-managing); a "No" response prompts an optional free-text comment.

## Study Phase 1: Tool development and retrospective validation

SCM Navigator was developed using 232 historical surgical cases across the three study specialties performed between December 2024 and April 2025. Each case was assigned a gold-standard SCM eligibility label (Affirmative/Maybe/Negative) through synchronous manual chart review by three study authors (two attending physicians and one medical student) with disagreements resolved through discussion until consensus was achieved. The operational determination made by the SCM attending during care was also recorded, enabling comparison among gold-standard labels, SCM Navigator classifications, and routine clinical adjudication.

Multiple LLM backbones—GPT4o-mini, GPT4o, o3-mini, GPT-4.1, and GPT-5 (all OpenAI)—were evaluated within the same agentic workflow, targeting 90% agreement with gold-standard labels to support deployment. OpenAI's o3-mini was selected for deployment based on performance and operational considerations (latency, cost, and availability within the production environment). Prompts were iteratively refined during development and fixed prior to deployment.

## Study Phase 2: Deployment and prospective evaluation

SCM Navigator was deployed with its outputs integrated into SHC's Epic instance as a System List displaying elective cases for the following day—labeled as Affirmative, Maybe, or Negative for SCM consultation (**Figure 1b**). The output was presented to SCM hospitalists as decision support: clinicians retained full discretion to confirm, decline, or override SCM Navigator's recommendations.

During deployment, SCM hospitalists were encouraged to document their triage decision for each reviewed case as **"Yes"** (anticipate co-managing) or **"No"** (do not anticipate co-managing), along with a brief reason when declining the tool's recommendation, via an EHR-embedded feedback form (Flow Sheet). Treating clinicians' in-workflow determinations were used as the reference standard for prospective evaluation.

### Physician Training

All SCM attending physicians were invited to participate, with 14 out of 16 participating. Training included online modules, handouts, and peer-to-peer demonstrations.

### Data Collection and Study Measures

SCM Navigator logged its classification (Affirmative/Maybe/Negative) and accompanying rationale, as well as treating clinicians' triage decisions and optional free-text reasons when declining a recommendation.

To support post-hoc error analysis, we reviewed discordant cases (i.e., SCM Navigator-Positive/Clinician "No," and SCM Navigator-Negative/Clinician "Yes"), where positive denotes either an Affirmative or Maybe flag. For putative false positives, clinicians' free-text feedback was manually reviewed and coded into prespecified categories (e.g., insufficient complexity, outpatient surgery, incompatible disposition, wrong primary service, outside-provider group designation) to characterize drivers of disagreement and distinguish model misclassification from workflow- or data-related factors. We also performed physician-led chart review of a subset of discordant cases, including all putative false-negatives and a purposive sample of putative false-positives representing cases stratified by whether SCM was subsequently consulted.

### Statistical Analysis

Data cleaning and performance metric computation were performed in Python[30]; statistical analyses and figures were generated in R using the tidyverse ecosystem[31]. Prospective performance was evaluated against clinician 'Yes/No' decisions, calculating sensitivity, specificity, and overall agreement (accuracy) with 95% CIs using 10,000 bootstrap replicates. Affirmative and Maybe recommendations were treated as positive triage decisions because both categories were intended to prompt human review. Cases without a recorded clinician decision were excluded.

## Results

### Study Phase 1: Tool development and retrospective validation

Development phase metrics are reported in **Figure 2**. In-workflow physician labeling was relatively error-prone, with an overall accuracy of 0.75 (95% CI 0.72–0.77) compared to gold-standard labels. While gpt-4o-mini performed worse than physicians with an overall accuracy of 0.71 (0.68-0.74), all other agents outperformed physicians: gpt-4o with an accuracy of 0.95 (0.92–0.97), o3-mini with an accuracy of 0.96 (0.93–0.98), gpt-4.1 with an accuracy of 0.94 (0.91–0.97), and gpt-5 with an accuracy of 0.94 (0.90–0.97).

Low accuracy among physicians was largely driven by missed eligible cases. Physicians showed moderate sensitivity (0.55, 95% CI 0.51–0.59), despite perfect specificity (1.0; 95% CI 1.0–1.0), meaning that nearly half of SCM-eligible patients were not flagged during routine review (**Figure 2b**). In contrast, the best-performing LLM achieved substantially higher sensitivities: 0.94 (95% CI 0.88–0.98) for gpt-4o, 0.91 (95% CI 0.84–0.96) for o3-mini, 0.86 (0.78–0.93) for gpt-4.1, and 0.91 (0.84-0.96) for gpt-5, while maintaining high specificities ranging from 0.95 to 0.99 (**Figure 2c**). Only gpt-4o-mini underperformed clinicians on sensitivity (0.23, 95% CI 0.17–0.29), though it exhibited perfect specificity.

Based on its competitive performance with other LLMs at lower cost, o3-mini was selected for deployment.

## Study Phase 2: Deployment and prospective evaluation

### Post-deployment performance

6,193 surgical cases were triaged by SCM Navigator in real-time since deployment on 9/22/2025, of which 1,582 (26%) were flagged as potentially appropriate for SCM: 1,429 (23%) were given a strong flag ("Affirmative") and 153 (2.5%) were given a weak flag ("Maybe"). Of the 6,193 triaged cases, 1,077 (17%) received physician feedback. Post-deployment performance metrics are calculated using physician adjudication as a reference standard and are reported in **Figure 3**. Label distributions ("Affirmative," "Maybe," or "Negative") between all cases and those that received feedback are shown in **Supplemental Table 1.**

SCM Navigator maintained high sensitivity (0.94, 95% CI 0.91–0.96), but specificity decreased to 0.74 (95% CI 0.71–0.77) compared to retrospective validation, reflecting a higher rate of false-positives during clinical use. Nonetheless, overall discrimination remained strong, with a class-balanced accuracy of 0.84 (95% CI 0.82–0.86) and an unbalanced accuracy of 0.79 (95% CI 0.77–0.82) (**Figure 3a-b**).

SCM Navigator displayed moderate PPV (0.58, 95% CI 0.54–0.63) and high NPV (0.97, 95% CI 0.95–0.98). As expected, the moderate PPV was partly driven by the low PPV among cases assigned the weak "Maybe" flag (class-specific PPV 0.23, 95% CI 0.12-0.35): the "Maybe" category was intentionally designed as a conservative, lower-confidence tier prompting optional review. In contrast, cases receiving the "Affirmative" strong flag had a substantially higher PPV

(class-specific PPV 0.63, 95% CI 0.58-0.67; **Figure 3c**). The high NPV indicates that Negative classifications were rarely overridden by physicians.

*Error Analysis*

### Feedback analysis

Manual review of physicians' free-text feedback among putative false-positive cases (n = 203) revealed three recurring patterns (**Figure 3d**).

First, one subset reflected limitations in upstream data reliability rather than errors in SCM Navigator's triage logic—including cases that were changed from inpatient to outpatient on the day of surgery (n=10; 4.9%) and cases where outside provider management was not documented (n=11; 5.4%).

Second, several cases reflected exclusion criteria that were not incorporated into SCM Navigator's triage logic during its design. These included incompatible level of care (n=30; 15%), often reflecting planned ICU-level post-operative care in high-risk neurosurgical cases, or SCM-incompatible primary service (n=22; 11%).

Third, the largest category—insufficient complexity (n=76; 37%)—covered patients who met some criteria but were not deemed to require co-management. To distinguish clinical judgment variability from SCM Navigator shortcomings, we performed chart reviews of discordant cases.

### In-depth chart review of select cases

We reviewed 30 putative false-positive cases and all 19 false-negative cases that received free-text feedback.

Among putative false-positive cases attributed to "insufficient complexity," we sampled 15 cases where SCM was subsequently consulted despite physician "No" (**Supplemental Table 2**) and 15 cases in which SCM was not subsequently consulted (**Supplemental Table 3**).

In the 15 cases with later SCM consultation (**Supplemental Table 2**), discordance most commonly reflected physician oversight of qualifying criteria (n=6; 40%) or evolving clinical complexity not apparent at triage (n=5, 33%). Less common drivers included patients undergoing multiple surgeries within a single admission who were already being followed by SCM (n=2; 13%), SCM-incompatible primary service (n=1; 6.7%), and level-of-care discrepancy (e.g. anticipated ICU-level post-operative care that limited SCM applicability until ICU de-escalation) (n=1; 6.7%).

In the 15 cases without later SCM consultation (**Supplemental Table 3**), 6/15 (40%) had comorbidities that met clear SCM criteria (e.g. atrial fibrillation, ankylosing spondylitis on biologic therapy). The remaining 9/15 (60%) reflected borderline complexity and practice-style variability, where patients could plausibly benefit from SCM but were deemed low-risk of exacerbation.

There were 19 putative false-negative cases (**Table 2**). The most common driver was noncomprehensive eligibility criteria (n=7; 37%). Additional drivers included physician adjudication errors (e.g. indicating "Yes" without realizing that the patient was managed by PAMF; n=3, 16%), complications arising after triage (n=2, 11%), and incomplete anesthesia preoperative documentation (n=2, 11%). A small number of reviewed cases reflected documentation/selection errors on the feedback form (n=3; 16%). Only two cases (11%) were attributed to LLM misclassification, in which clear rubric-qualifying criteria were documented but SCM Navigator returned "Negative". In total, we estimate the true-negative rate in the putative false-negative cohort to be 42% (by aggregating physician errors in adjudication, unidentified PAMF patients, and selection errors; n=8) at minimum.

## Discussion

This prospective implementation study evaluates an LLM-powered, EHR-integrated surgical triaging tool deployed at a large academic medical center. SCM Navigator achieved high sensitivity (94%), demonstrating that an LLM-based agentic workflow can accurately augment clinical triaging protocols. Specificity (78%) was lower to intentionally allow clinical judgment in ambiguous cases. In this context, higher sensitivity and lower specificity may optimize patient safety.

Safety played a paramount consideration in the design and implementation of the tool. Of the 1,077 cases with direct physician feedback, 182 cases differed in SCM Navigator and physician determination (163 false-positive, 19 false-negative). We considered false-negative cases as particularly high-risk if a patient were to be missed by SCM Navigator and not receive appropriate SCM care. Most false-negatives were attributable to noncomprehensive clinical criteria, structural workflows such as outside medical group coverage, and medical complications unrelated to the patient's baseline medical comorbidities—factors addressable through additional SCM Navigator configuration. Only 11% were attributable to LLM error.

We sought to understand where SCM Navigator could be further optimized. The largest category of false-positives was insufficient complexity (76 cases), perhaps reflecting variations in clinical practice patterns among participating physicians attributable to differential departmental staffing support, familiarity with patient populations of specific services (every SCM hospitalist is assigned a home surgical service), years of practice experience, and personal practice styles. Structural misclassification was the next largest category (e.g. patients managed by an outside group not labeled as such in the EHR, high-risk surgeries automatically routed to postoperative ICU rather than floor-level care). These suggest that SCM Navigator errors were driven more by incomplete criteria formation, incomplete workflow programming, and lack of practice standardization—all addressable with additional structured checks.

Notably, some false-positives reflected provider error and practice variability, wherein SCM Navigator flagged patients who clearly met SCM eligibility but were not identified by physicians, or ambiguous patients who subsequently developed SCM need. Manual review of 15 randomly selected false-positive cases for whom SCM was not later consulted demonstrated that 40% would have confidently met SCM eligibility, suggesting instances wherein SCM Navigator outperformed providers in triaging.

Our study exemplifies a larger trend of LLMs delivering the potential to augment clinical protocols in real-world settings.[32–34] We hypothesize that SCM Navigator standardizes disparate practice patterns while flagging patients who may not have been identified otherwise. To the extent that standardized, evidence-based identification leads to more consistent and timely surgical co-management, both patients and providers stand to benefit. With further refinement of the clinical criteria in its prompt, SCM Navigator may be optimized to reduce its false-positive and false-negative rate, potentially maturing from augmentation to automation of high-risk patient identification.

Implementation challenges included stakeholder engagement, integration into existing workflows, feedback consensus, and the time commitment required for ongoing evaluation. Where provider decision-making is idiosyncratic, technology will be challenged to match individual provider practice patterns. These lessons may inform other health systems deploying agentic AI workflows. Multi-stakeholder buy-in—particularly from end-users—is paramount for tool adoption and use. Ongoing feedback must be solicited and adjudicated across stakeholder perspectives. Most importantly, implementations must be designed to test and assure safety before any novel deployment.

The success of this LLM-driven triage hinged on both evidence-based clinical criteria and organization-specific practice patterns, with discrepancies partly reflecting insufficient incorporation of the latter (e.g. outpatient vs inpatient surgeries, postoperative care at non-SCM locations, and patients managed by other physician groups). This highlights that effective automation within healthcare settings must reflect not only clinical guidelines but also institutional practice patterns.

Limitations of this study include single-center design and moderate feedback rate (18%) from physicians. Accuracy evaluation was performed against physician feedback within SCM Navigator rather than encounter billing. Only a subset of triaged patients received retrospective manual review. Enthusiastic physicians may have disproportionately contributed more feedback, potentially overrepresenting particular practice patterns. Nonetheless, we confirmed that many physicians who did not directly provide feedback still utilized SCM Navigator as an initial screen.

Future investigations may evaluate the clinical impact of tool utilization on patient outcomes, operational throughput, time spent within the EHR by clinicians, and total cost of care. A retrospective clinical trial may be conducted to evaluate the effect of SCM Navigator on postoperative complications, rapid responses, length of stay, and total cost of care before and after tool implementation. Further evaluation should be conducted on SCM Navigator-physician discrepant determinations, clinician feedback on workflow processes, as well as clinician satisfaction with tool output, use, and interface, and time spent within the EHR.

## Conclusion

A human-in-the-loop, LLM-based, EHR-integrated agentic workflow can triage surgical patients for SCM consultation with high sensitivity and moderate specificity. SCM Navigator was associated with zero patient safety events after six months of clinical deployment, with most analyzed discrepant cases attributable to suboptimal clinical criteria or structural nuances addressable through further programming. AI-based clinical workflow tools have the potential to augment and eventually automate existing clinical workflows. Successful implementation requires not only technical accuracy and safety guardrails, but stakeholder alignment from builders, end-user providers, and operational leadership.

## Acknowledgements

We would like to express particular gratitude the Surgical Co-Management section of the Stanford University Division of Hospital Medicine for their support and collaboration in evaluating and refining the SCM Navigator, the Division of Hospital Medicine for their support of this collaboration, and the Stanford Technology and Digital Services team, in particular the Data Science team, for their pioneering work in the development of ChatEHR tool as well as unwavering technical and visionary support in the conception, development, and ongoing evaluation of the SCM Navigator.

Jane Wang is a General Partner in MedMountain Ventures, LLC.

Jason Hom is an advisor and option holder for Cognita Imaging and Radiology Partners, receives support from ARPA-H (1AY2AX000045-01), the NIH (5U01NS134358-07), Stanford HAI, and the Gordon and Betty Moore Foundation.

Dr. Jonathan Chen is the Co-founder of Reaction Explorer LLC that develops and licenses organic chemistry education software, is a paid medical expert witness fees from Elite Experts, and has been paid one-time honoraria or travel expenses for invited presentations by insitro, General Reinsurance Corporation, AASCIF, and other industry conferences, academic institutions, and health systems. He has received research funding support in part by the NIH/National Institute of Allergy and Infectious Diseases (1R01AI17812101), the NIH-NCATS-Clinical & Translational Science Award (UM1TR004921), Stanford Bio-X Interdisciplinary Initiatives Seed Grants Program (IIP) [R12] [JHC], the NIH/Center for Undiagnosed Diseases at Stanford (U01 NS134358), Stanford RAISE Health Seed Grant 2024, Josiah Macy Jr. Foundation (AI in Medical Education), and Stanford CARE AI Scholar Fellowship.

Timothy Keyes, April Liang, Stephen Ma, Jason Shen, Jerry Liu, Nerissa Ambers, Abby Pandya, Rita Pandya, Natasha Steele, and Kevin Schulman do not have disclosures relevant to this study.

# Figures

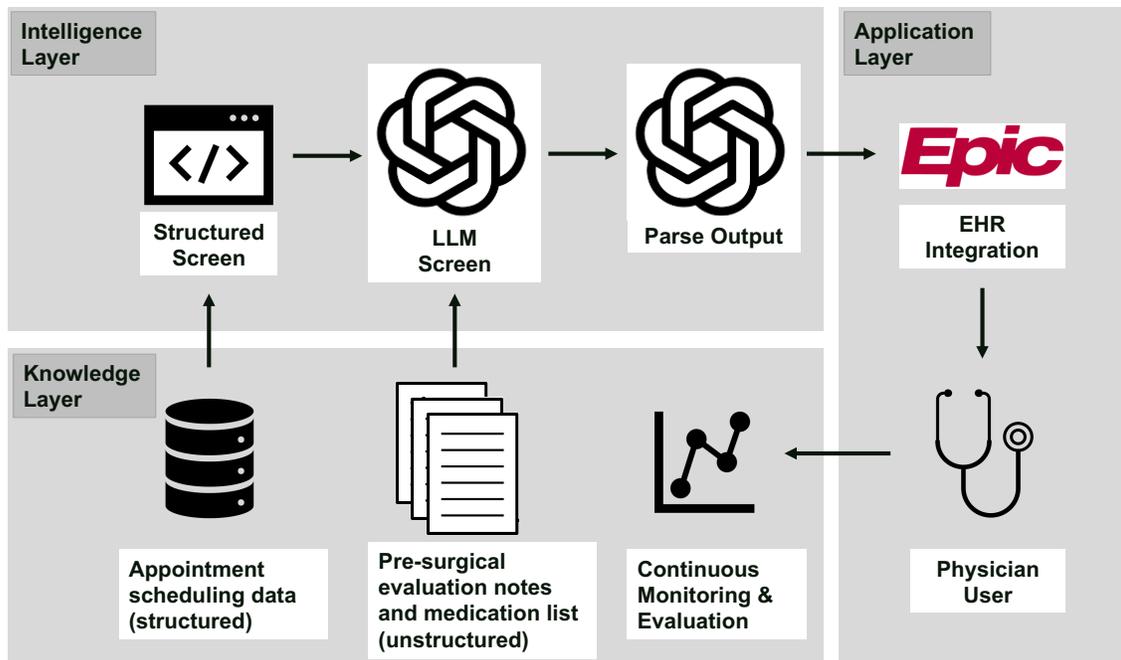

**Figure 1 – Architecture and user-facing workflow for SCM Navigator.**

**(A)** Conceptual diagram of SCM Navigator, organized using Li et al.'s four-layer care transformation framework for health care AI[35]. The *Knowledge Layer* (data) combines structured scheduling data with unstructured pre-operative documentation (pre-operative evaluation note and medication list). The *Intelligence Layer* (LLM and agentic workflow) applies two sequential LLM steps—(i) an LLM "triage" that generates a case-level eligibility determination for surgical co-management (SCM) with supporting evidence and (ii) a parsing step that converts the free-text output into structured fields for logging and downstream use by the Application Layer. The *Application Layer* (EHR integration and user-interface) integrates outputs into the electronic health record (EHR), enabling review and action by clinicians. Continuous monitoring and evaluation of physician-provided, in-workflow feedback are incorporated to support prospective performance measurement and error analysis.

**(B)** Screenshot of SCM Navigator output as displayed to SCM hospitalists within an Epic Hyperspace, showing next-day elective cases with model-generated triage results to support clinician review and triage (*Workflow Layer*).

## A

### Performance Metrics by Classification Method

|  | In-workflow Physician | gpt-4o-mini | gpt-4o | o3-mini | gpt-4.1 | gpt-5 |
|---|---|---|---|---|---|---|
| Accuracy | 0.75 [0.72, 0.77] | 0.71 [0.68, 0.74] | 0.95 [0.92, 0.97] | **0.96 [0.93, 0.98]** | 0.94 [0.91, 0.97] | 0.94 [0.90, 0.97] |
| Balanced Accuracy | 0.78 [0.75, 0.80] | 0.62 [0.59, 0.65] | 0.95 [0.92, 0.97] | **0.95 [0.91, 0.98]** | 0.92 [0.88, 0.96] | 0.93 [0.89, 0.96] |
| NPV | 0.63 [0.59, 0.66] | 0.68 [0.65, 0.71] | 0.96 [0.93, 0.99] | **0.95 [0.91, 0.98]** | 0.92 [0.88, 0.96] | 0.95 [0.91, 0.98] |
| PPV | 1.0 [1.0, 1.0] | 1.0 [1.0, 1.0] | 0.93 [0.88, 0.98] | **0.97 [0.94, 1.0]** | 0.97 [0.93, 1.0] | 0.92 [0.85, 0.97] |
| Sensitivity | 0.55 [0.51, 0.59] | 0.23 [0.17, 0.30] | 0.94 [0.89, 0.98] | **0.91 [0.84, 0.96]** | 0.86 [0.78, 0.93] | 0.91 [0.84, 0.96] |
| Specificity | 1.0 [1.0, 1.0] | 1.0 [1.0, 1.0] | 0.96 [0.93, 0.99] | **0.99 [0.96, 1.0]** | 0.99 [0.97, 1.0] | 0.95 [0.92, 0.99] |

## B

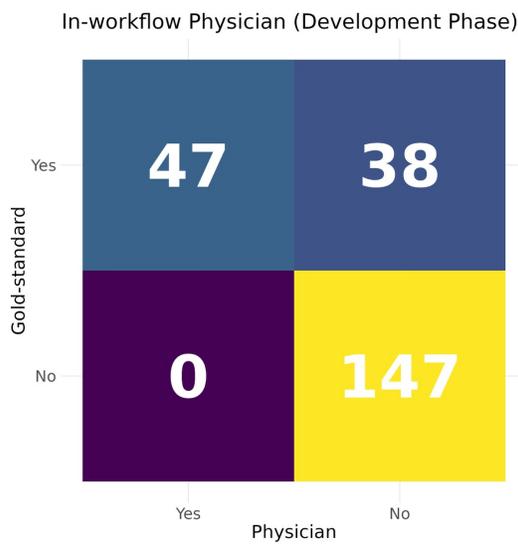

## C

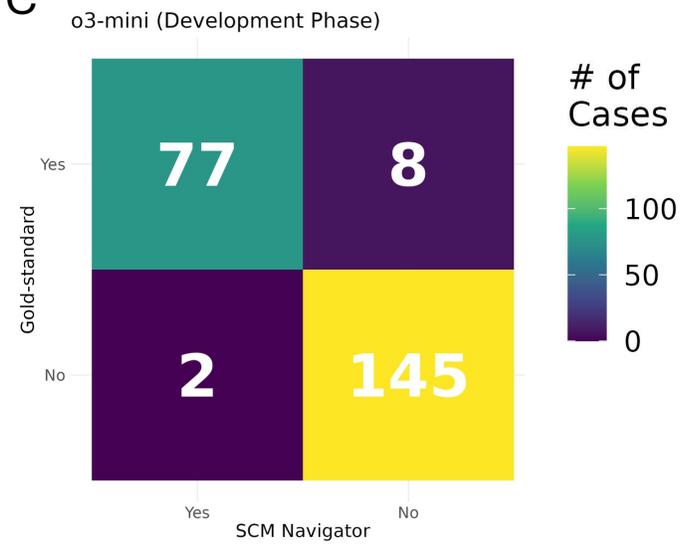

**Figure 2 – Retrospective (pre-deployment) performance of SCM Navigator across candidate LLM backbones.**

**(A)** Performance of SCM Navigator across candidate LLM backbones in the development dataset (N=232), benchmarked against the in-workflow physician triage and gold-standard eligibility labels derived from consensus chart review. Metrics are shown with 95% confidence intervals from bootstrap resampling (gray, bracketed); the deployed o3-mini configuration is bolded.

**(B - C)** Confusion matrices comparing triage decisions against gold-standard labels for (B) the in-workflow physician triage and (C) the o3-mini–powered SCM Navigator selected for deployment. The o3-mini SCM Navigator substantially reduced false-negatives (gold-standard eligible cases triaged as not eligible; top-right cells) relative to in-workflow physician triage, with only a small increase in false-positives (bottom-left cells) in the development dataset.

Across panels, SCM Navigator outputs are displayed as a binary triage (collapsing Affirmative/Maybe into a single positive class) to align with the in-workflow physician adjudication, which was recorded as a binary Yes/No.

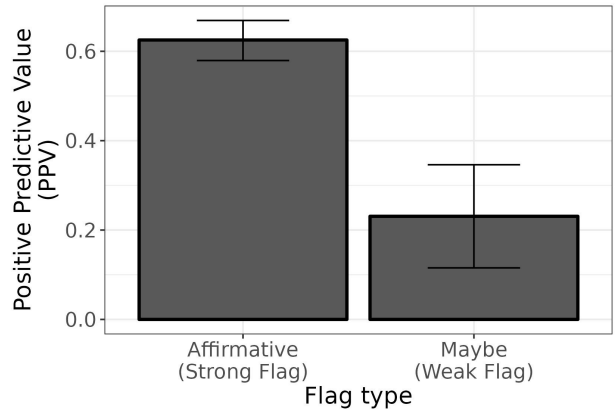
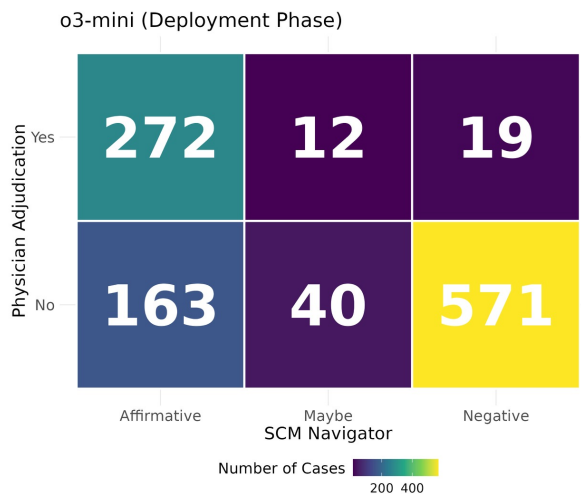
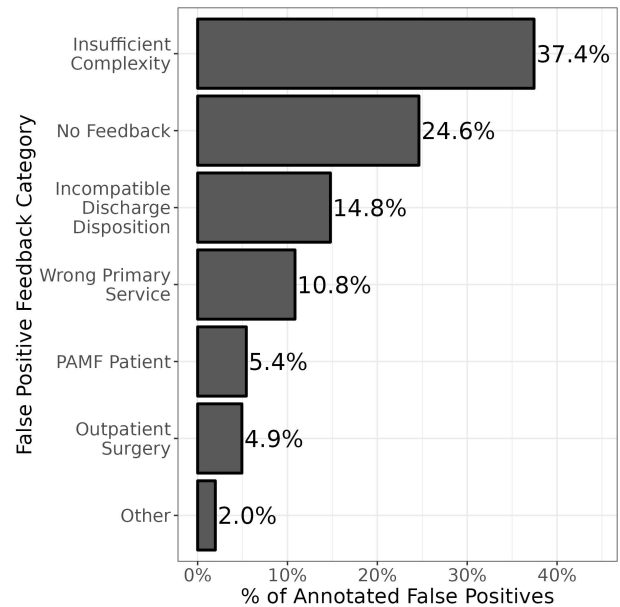

**Figure 3 – Prospective (post-deployment) performance and error categorization of SCM Navigator outputs**.

**(A)** Post-deployment performance metrics for the o3-mini–powered SCM Navigator, calculated among cases with in-workflow treating-physician feedback during routine clinical use. Metrics are reported with 95% confidence intervals estimated via bootstrap resampling.

**(B)** Confusion matrix comparing SCM Navigator triage output against treating-physician Yes/No determinations (collected via in-workflow feedback) during the deployment phase of the study. Most of SCM Navigator's post-deployment errors were identified as false-positives: patients identified as potentially appropriate for SCM by SCM Navigator, but not appropriate for SCM by the treating SCM physician.

**(C)** Positive predictive value (PPV) stratified by SCM Navigator flag type among cases with physician feedback. The strong "Affirmative" flag demonstrated substantially higher PPV than the weak "Maybe" flag, consistent with "Maybe" functioning as a lower-confidence tier intended to prompt optional clinician review. Error bars indicate 95% bootstrap confidence intervals.

**(D)** Distribution of false-positive feedback themes among cases in which SCM Navigator flagged a case as potentially SCM-appropriate but the treating physician declined consultation. Categories were derived by manual review of physicians' free-text reasons and grouped into recurring drivers of disagreement. The most common theme was *insufficient complexity* (i.e., patients meeting prompt-specified risk factors but judged not to warrant co-management based on SCM clinicians' clinical judgment), followed by operational/workflow exclusions (e.g., incompatible discharge disposition, wrong primary service, outpatient surgery, or outside-provider group designation), suggesting that a substantial portion of post-deployment "errors" reflected workflow or judgment variability rather than classification errors made by SCM Navigator.

# Tables

**Table 1- Clinical criteria used to formulate SCM Navigator.**

| Criterion | Reference |
|---|---|
| History stroke or cerebrovascular accident (CVA) | [28,36] |
| History congestive heart failure (HFrEF or HFpEF) | [28,37] |
| History CAD, STEMI, NSTEMI, CABG, PCI | [28] and shared clinical expertise of SCM faculty |
| History of cardiac arrhythmia | [28,38] |
| History of diabetes (Type 1 or 2) | [28] and shared clinical expertise of SCM faculty |
| History of hypertensive | [28] and shared clinical expertise of SCM faculty |
| History of dementia or cognitive impairment | [28,29] |
| History of Crohn's or ulcerative colitis | [39] and shared clinical expertise of SCM faculty |
| Adrenal insufficiency or chronic steroid use prednisone or other steroids, or chronic steroid use | Shared clinical expertise of SCM faculty |
| History of COPD and on controller medications | [28] |
| History of asthma with exacerbations in the last 12 months | Shared clinical expertise of SCM faculty |
| History of chronic kidney disease (CKD) or dialysis use | [28,36] and shared clinical expertise of SCM faculty |
| Obstructive sleep apnea (OSA) | [28] |
| Over 2 medical diagnoses | [28,37] |
| Active medication list includes biologics | Shared clinical expertise of SCM faculty |
| Active medication list includes cancer treatment | Shared clinical expertise of SCM faculty |

**Table 2 - Post-hoc manual physician review of putative false-negative cases.**

| Classification | Number of Cases (N=19) | Example |
|---|---|---|
| Clinical criteria noncomprehensive | 7 | Patient had diagnosis of active substance use, but this was not included in SCM Navigator clinical triage criteria |
| Preoperative or postoperative medical complication | 2 | Patient at baseline would not have required SCM services, but developed postoperative hypotension requiring vasopressors. |
| Uncomprehensive or lack of anesthesia preoperative note | 2 | Anesthesia preoperative note did not include critical comorbidities that would have screened the patient in |
| PAMF patient | 3 | Patient was designated as a PAMF member by SCM Navigator, but treating physician may not have seen the PAMF label, resulting in an erroneous positive label |
| LLM error | 2 | LLM incorrectly categorized patient as not requiring SCM services when anesthesia preop note clearly documents clinical criteria that would meet eligibility |
| Other | 3 | SCM provider incorrectly clicked "Yes" that the patient needed to be seen even when they did not meet any known criteria. Possible point-and-click error. |

# Supplemental Materials

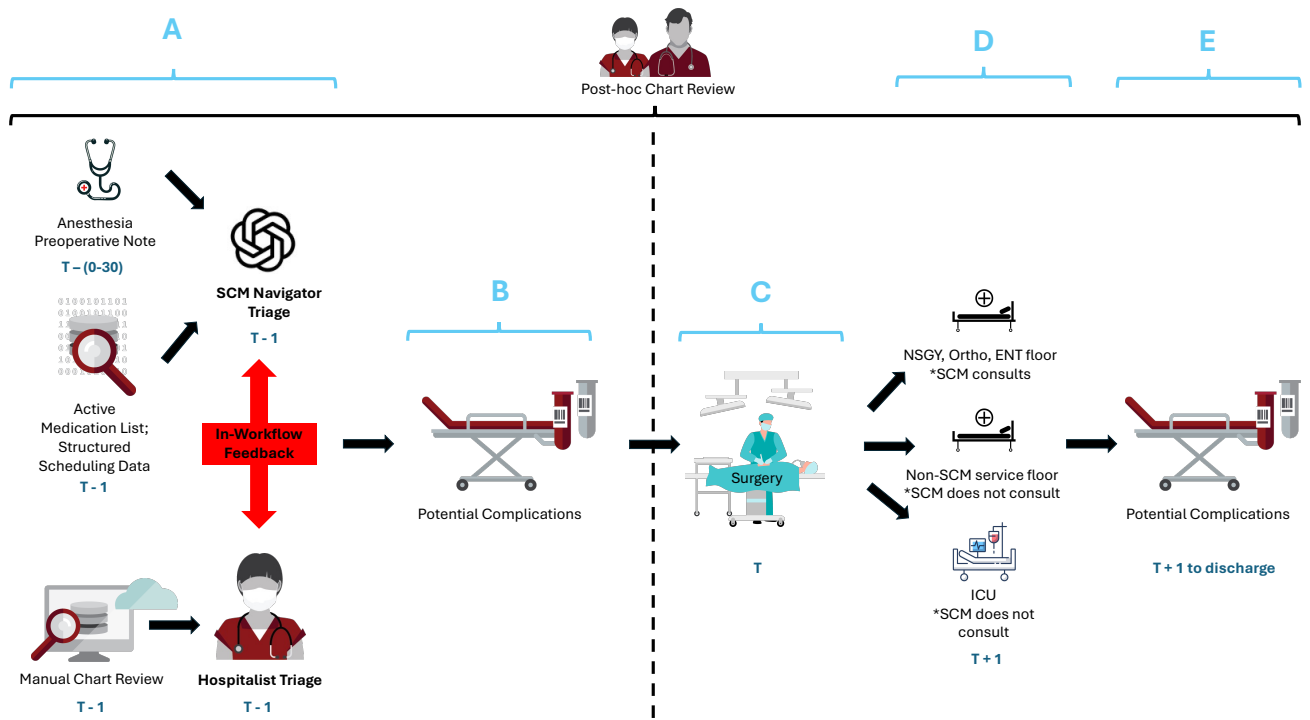

**Supplemental Figure 1 – End-to-end clinical workflow for SCM Triage and perioperative co-management**

**(A) Pre-Operative AI and Physician Screening.** Beginning up to 30 days before surgery (T-(0-30)), the anesthesia preoperative note and medication list for elective patient surgeries are documented in the EHR. On the day before surgery (T-1), these data are fed into SCM Navigator, which performs an in-workflow risk scoring protocol for the following day's surgical cases. The treating SCM hospitalist then reviews the output from the SCM Navigator, compares it against their own manual chart review, and determines whether a patient should be co-managed by SCM.

**(B) Potential Pre-Operative Complications.** For patients admitted from the emergency department or outside hospitals, additional comorbidities may be accounted for upon further assessment. These clinical developments are typically not visible to SCM Navigator.

**(C) Surgical Intervention**. On day T, the patient receives surgical intervention.

**(D) Post-Operative Destination.** Post-operatively, on the same day of the surgery (T) or the day after (T+1), the patient can be transitioned to the Neurosurgery (NSGY), Orthopedics (Ortho), or

Otolaryngology (ENT) floors, on which SCM is a consulting service. SCM is not a consulting service on other hospital floors or in the intensive care unit (ICU).

**(E) Potential Post-Operative Complications.** Surgical or medical complications can arise after the surgery up through the time of discharge (T to discharge), at which time SCM may become involved in a patient's care.

The dashed vertical line denotes the information cutoff for post-hoc chart review: reviewers based their assessments solely on clinical information available before this timepoint, mirroring the data available to SCM Navigator and physicians at the time of triage.

**Supplemental Table 1: Distribution of cases by SCM Navigator recommendation against physician feedback and agreement**

|  | All Cases | Cases With Feedback | Feedback Rates by SCM Navigator Class | Physician Agreement |
|---|---|---|---|---|
| **Total N** | 6193 (100%) | 1077 (100%) | 1077 out of 6193 (17%) | 855 out of 1077 (79%) |
| **Affirmative** | 1429 out of 6193 (23%) | 435 out of 1077 (40.%) | 435 out of 1429 (30.%) | 272 out of 435 (63%) |
| **Maybe** | 153 out of 6193 (2.5%) | 52 out of 1077 (4.8%) | 52 out of 153 (34%) | 12 out of 52 (23%) |
| **Negative** | 4611 out of 6193 (74%) | 590 out of 1077 (55%) | 590 out of 4611 (13%) | 571 out of 590 (97%) |

**Supplemental Table 2 - Post-hoc manual physician review of putative false-positive cases subsequently consulted by SCM.**

| Classification | Number of Cases (N=15) | Example |
|---|---|---|
| Physician error | 6 | Provider failed to identify key clinical criteria (ex. atrial fibrillation, diabetes requiring insulin, etc) qualifying patient for SCM services; surgical service subsequently consulted SCM to co-follow |
| Physician practice style | 5 | Patient with advanced age, HTN, HLD, prediabetes that met clinical triage guidelines, but provider chose not to see. Patient later developed oxygen requirement, and surgical team requested SCM consultation |
| Patient already being followed by SCM | 2 | Patient had multiple surgical procedures completed in same admission, was followed by SCM throughout. |
| Patient already on non-SCM primary service | 1 | Patient initially admitted to medicine service, then to surgical service for procedure, transferred to ICU postoperative, then downgraded to floor-level care (SCM consulted then) |
| Discharge disposition | 1 | Provider correctly judged that patient would be transferred to ICU postoperatively |

**Supplemental Table 3 – Post-hoc manual physician review of putative false-positive cases not subsequently consulted by SCM.**

| Classification | Number of Cases (N=15) | Example |
|---|---|---|
| Physician error | 6 | Patient with sufficient comorbidities that could result in life-threatening exacerbation or complication, which would merit SCM consultation. |
| Physician practice style | 9 | Patient with certain number and type of medical comorbidities that may or may not have exacerbated, and that certain surgical teams are comfortable managing independent of SCM. |

# Appendix 1 – SCM Navigator Prompts.

**System Prompt**

*Purpose.*

In addition to the task-specific prompts below, we supplied a system prompt to establish general behavior expectations for the LLM (e.g., role framing and constraints). This system prompt was provided as the system-level instruction in our LangChain workflow and remained constant across all cases.

*Prompt text.*

```
# Role

You are a medical language assistant with exceptional attention to
detail. Your primary responsibility is to carefully review electronic
health records (EHRs) and determine whether patients meet specific
eligibility criteria for medical interventions.

# Approach

      You meticulously analyze each patient chart, verifying all
      information against the provided eligibility criteria.
      You only make determinations that are explicitly supported by the
      data in the patient's chart.
      You do not make assumptions, infer missing details, or introduce
      external knowledge.
      If a criterion is not directly supported by the information
      provided, you should provide a negative response, but note that
      there may have been insufficient information.

Always verify your conclusions before finalizing your response.
```

**Classification Prompt**

*Purpose.*

This prompt is used to elicit an initial eligibility determination for the Surgical Co-management program from a patient's Anesthesia Pre-procedure Evaluation note and medication list. The model is instructed to (1) apply a fixed set of eligibility criteria, (2) return a categorical label ("Affirmative", "Maybe", "Negative") and (3) justify the label using explicit supporting text from the note and the numbered criterion/criteria. This prompt is provided to the model as the user-facing task instruction within an LLM workflow implemented using LangChain[40].

*Prompt text.*

```
# Task

This patient is scheduled to have surgery.

We need your help determining if the patient meets the requirements
for a special medical program based on the eligibility criteria
described below.

# Eligibility ("Affirmative") criteria

## Medical History

A surgical patient is appropriate for the program if their past
medical history includes any of the following:

1) History of cerebrovascular accidents (CVA) or transient ischemic
attacks (TIA)
2) History of chronic metabolic syndrome (CMY), heart failure with
preserved ejection fraction (HFpEF), or heart failure with reduced
ejection fraction (HFrEF)
3) History of coronary artery disease (CAD), including STEMI, NSTEMI,
CABG, or PCI
4) Diabetes
5) Hypertension
6) Dementia (**do not** confuse this with minor psychiatric
conditions, like anxiety, depression, or insomnia)
7) A diagnosis of Crohn's disease or Ulcerative Colitis
8) A history of any serious arrhythmias (e.g. atrial fibrillation,
supraventricular tachycardia). Note that having premature ventricular
contractions (PVCs) alone does **not** meet this criterion.

## Current medical problems

A surgical patient is appropriate for the program if their current
medical needs include any of the following:
```

9) The patient is **currently** receiving ANY biologic treatments (like monoclonal antibodies)
10) The patient is **currently** taking medications (such as infusion medications) for any cancer. A **previous** medical history of cancer does NOT meet this criterion and should be ignored. Only **active** medications for cancer meet this criterion.

# "Maybe" criteria

A surgery patient is a "maybe" for the special medical program if they do not meet any of the criteria mentioned above, but they meet one of the following:

11) The patient has an active malignancy (of any kind). Note that having a **past** history of cancer (that is now in remission) alone does **not** meet this criterion.
12) The patient has a history of hypertension that is specifically described as "well-controlled," "diagnosed many years ago without complications," or similar.
13) The patient has three or more medical comorbidities not described above, particularly the following: obstructive sleep apnea (OSA), an autoimmune disorder, adrenal insufficiency or chronic steroid use, chronic obstructive pulmonary disease (COPD), or asthma **with exacerbation in the past 12 months**. These comorbidities should be relatively serious medical conditions (e.g. **not** anxiety, insomnia, depression, a minor injury, or similar).
14) The patient has a cancer history that requires them to continue to take medications (e.g. tamoxifen or other aromatase inhibitors) or may increase thrombotic risk.

# Patient's Medical Documentation

Now, please read through the clinical documentation written by the patient's care team. Pay particular attention to whether the documentation mentions any of the criteria mentioned above.

{documentation}

# Final request

If the patient meets one or more of the Affirmative eligibility criteria, provide a string response of "Affirmative". If they do not meet any of the Affirmative Eligibility criteria but **do** meet one or more of the "Maybe" criteria, provide a string response of "Maybe".
Otherwise, respond with a string response of "Negative".

In addition to this string response, include a brief explanation of
why that answer is correct. The explanation must provide a summary of
the patient's problems (as mentioned in the documentation), a direct
quote of the patient note indicating where the criteria are mentioned,
**and** the number of the criterion or criteria that are met.

Remember, you must strictly evaluate the criteria above - do not infer
the patient's medical needs or form an interpretation beyond the
explicit criteria that have been provided.

**Parsing Prompt**

*Purpose.*

The model's initial response to the Classification Prompt is free text. To facilitate consistent downstream analysis and evaluation, we apply a second LLM step that extracts the classification label and explanation into a standardized structured format.

The {raw_response} field below contains the raw text output produced by the Classification Prompt stage (i.e., the predicted label and explanation).

*Prompt text.*

```
You are a structured extraction layer.

Convert the passage below into the `ScreeningResult` schema.

Rules:
- Use only the information in the passage.
- Do not add new facts.
- Normalize the classification to exactly one of:
  - Affirmative
  - Maybe
  - Negative
- If the passage mentions that the documentation was insufficient,
missing, or ambiguous, be sure to explain the uncertainty in
`explanation`. When this happens, urge the user to review the patient
manually in the first sentence of the explanation.
```

```
- Return the structured result only.

Passage:
{raw_response}
```

*Structured output schema (Pydantic)*

The structured output contains two fields:
- *Classification* — one of "Affirmative," "Maybe," or "Negative." Defaults to "Maybe" if parsing fails.
- *Explanation* — a free-text string summarizing the relevant patient evidence, the criterion or criteria met, any supporting quoted text from the documentation, and any noted uncertainty or ambiguity.

**Data Sharing Statement:** The data used in this study are derived from live electronic health record systems at Stanford Health Care and contain protected health information. These data cannot be made publicly available due to patient privacy regulations and institutional policies governing the use of clinical data. Requests for analytic methods or additional information about the study may be directed to the corresponding author.